\documentstyle[12pt]{article}

\catcode`\@=11
\def\marginnote#1{}
\newcount\hour
\newcount\minute
\newtoks\amorpm
\hour=\time\divide\hour by60
\minute=\time{\multiply\hour by60 \global\advance\minute by-\hour}
\edef\standardtime{{\ifnum\hour<12 \global\amorpm={am}%
        \else\global\amorpm={pm}\advance\hour by-12 \fi
        \ifnum\hour=0 \hour=12 \fi
        \number\hour:\ifnum\minute<10 0\fi\number\minute\the\amorpm}}
\edef\militarytime{\number\hour:\ifnum\minute<10 0\fi\number\minute}
\def\draftlabel#1{{\@bsphack\if@filesw {\let\thepage\relax
   \xdef\@gtempa{\write\@auxout{\string
      \newlabel{#1}{{\@currentlabel}{\thepage}}}}}\@gtempa
   \if@nobreak \ifvmode\nobreak\fi\fi\fi\@esphack}
        \gdef\@eqnlabel{#1}}
\def\@eqnlabel{}
\def\@vacuum{}
\def\draftmarginnote#1{\marginpar{\raggedright\scriptsize\tt#1}}
\def\draft{\oddsidemargin -.5truein
        \def\@oddfoot{\sl preliminary draft \hfil
        \rm\thepage\hfil\sl\today\quad\militarytime}
        \let\@evenfoot\@oddfoot \overfullrule 3pt
        \let\label=\draftlabel
        \let\marginnote=\draftmarginnote
   \def\@eqnnum{(\theequation)\rlap{\kern\marginparsep\tt\@eqnlabel}%
\global\let\@eqnlabel\@vacuum}  }

\def\preprint{\twocolumn\sloppy\flushbottom\parindent 1em
        \leftmargini 2em\leftmarginv .5em\leftmarginvi .5em
        \oddsidemargin -.5in    \evensidemargin -.5in
        \columnsep 15mm \footheight 0pt
        \textwidth 250mmin      \topmargin  -.4in
        \headheight 12pt \topskip .4in
        \textheight 175mm
        \footskip 0pt
        \def\@oddhead{\thepage\hfil\addtocounter{page}{1}\thepage}
        \let\@evenhead\@oddhead \def\@oddfoot{} \def\@evenfoot{} }

\def\titlepage{\@restonecolfalse\if@twocolumn\@restonecoltrue\onecolumn
     \else \newpage \fi \thispagestyle{empty}\c@page\z@ 
        \def\thefootnote{\fnsymbol{footnote}} }

\def\endtitlepage{\if@restonecol\twocolumn \else  \fi
        \def\thefootnote{\arabic{footnote}}
        \setcounter{footnote}{0}}  

\catcode`@=12
\relax
\def\beq{\begin{equation}}
\def\eeq{\end{equation}}

\def\Im{\mathop{\rm Im}}
\def\NP#1#2#3{Nucl. Phys. \underline{#1} (19#2) #3}
\def\ov{\overline}
\def\PL#1#2#3{Phys. Lett. \underline{#1} (19#2) #3}
\def\PR#1#2#3{Phys. Rev. \underline{#1} (19#2) #3}
\def\PRL#1#2#3{Phys. Rev. Lett. \underline{#1} (19#2) #3}
\def\Re{\mathop{\rm Re}}

\def\crbig{\\\noalign{\vspace {3mm}}}

\def\R{\bf R}
\def\Z{\bf Z}
\relax

\begin{document}
\topmargin-2.4cm
\renewcommand{\theequation}{\thesection.\arabic{equation}}
\begin{titlepage}
\begin{flushright}
NEIP--00--005 \\
hep--th/0003078 \\
March 2000
\end{flushright}
\vspace{2.4cm}

\begin{center}
{\Large\bf On the Effective N=1 Supergravity 
of M-Theory$^\star$}
\vskip .8in
{\bf J.-P. Derendinger and R. Sauser}$^{\dagger}$
\vskip .1in
{\it Institute of Physics,
University of Neuch\^atel \\
CH--2000 Neuch\^atel, Switzerland}
\end{center}

\vspace{1.9cm}

\begin{center}
{\bf Abstract}
\end{center}
\begin{quote}
In the low-energy limit, M-theory compactified on $S^1/{\bf Z}_2$ is 
formulated in terms of Bianchi identities with sources localized
at orbifold singularities and anomaly-cancelling counterterms to 
the Wilson effective Lagrangian. Compactifying to four dimensions
on a Calabi-Yau space leads to $N=1$ local supersymmetry. 
We derive a formulation of the effective supergravity which explicitly 
relates four-dimensional supergravity multiplets and field equations
with these fundamental M-theory aspects. This formulation proves 
convenient for the introduction in the effective supergravity
of non-perturbative M-theory contributions. 
It also applies to the universal sector of generic
compactifications with $N=1$ supersymmetry. 
\end{quote}

\vfill
\begin{flushleft}
\rule{8.1cm}{0.2mm}\\
$^{\star}$
{\small Research supported in part by
the Swiss National Science Foundation.} \\
$^{\dagger}$ {\small\tt jean-pierre.derendinger, roger.sauser@iph.unine.ch}
\end{flushleft}

\end{titlepage}
\setcounter{footnote}{0}
\setcounter{page}{0}
\setlength{\baselineskip}{.7cm}
\newpage 
\newpage

\section{Introduction}\label{secintro}
\setcounter{equation}{0}

M-theory compactified on
\beq
\label{O7is}
O_7 = X_6\times S^1/{\bf Z}_2\,,
\eeq
where $X_6$ is a Calabi-Yau three-fold, leads to a four-dimensional 
theory with $N=1$ local supersymmetry. At present, our 
knowledge of the resulting effective supergravity  
is based on the few aspects of M-theory 
which are quantitatively understood and on the small-orbifold 
limit\footnote{The limit in which the length 
of the $S^1/{\bf Z}_2$ orbifold direction 
is small compared to the size of the Calabi-Yau manifold.} which is 
the perturbative heterotic $E_8\times E_8$ string compactified 
on the Calabi-Yau space $X_6$. 
In the low-energy limit, M-theory information can be 
organized as an expansion in powers of the eleven-dimensional
gravitational constant $\kappa_{11}$ \cite{HW1, HW2}. 
The lowest order $\kappa^{-2}_{11}$ is eleven-dimensional
supergravity \cite{CJS}. In the case of a compactification on 
$S^1/{\bf Z}_2$ only, 
the next orders in $\kappa_{11}$ are known to include orbifold 
plane contributions (super-Yang-Mills terms)
as well as gauge and gravitational anomaly-cancelling terms 
\cite{HW1, W, HW2}. 

Similarly, the effective four-dimensional 
supergravity\footnote{Which is a low-energy description.} can 
be formulated as an expansion in the four-dimensional gravitational 
constant $\kappa$, even if a more common choice suggested 
by string theory is to use the dilaton as 
expansion parameter. The two options coincide provided the 
appropriate supersymmetric description of the dilaton is adopted in
a Wilson effective Lagrangian formulation.
The lowest order $\kappa^{-2}$ is the $O_7$ {\it truncation} of 
eleven-dimensional supergravity. The next order includes 
super-Yang-Mills, charged matter kinetic and superpotential
contributions. Then come sigma-model anomaly-cancelling 
terms contributing in particular to gauge threshold corrections.  
These first corrections to the low-energy limit of compactified 
M-theory are identical to those obtained from heterotic 
compactifications on Calabi-Yau. This is certainly 
expected since the information content is identical. 
The literature gives a detailed description of
these results, with particular attention paid to the 
`strong-coupling' heterotic limit in which the size 
of the Calabi-Yau space is smaller than the orbifold 
length, supersymmetry breaking by gaugino condensation and 
non-standard embeddings \cite{ref1}--\cite{nonstandard}. 

Our goal in this paper is to provide a derivation of the
effective supergravity which explicitly relates four-dimensional 
supergravity statements with M-theory aspects like Bianchi
identities modified at singularities and anomaly cancellation. 
We reformulate these basic facts 
of M-theory on $O_7$ directly in terms of four-dimensional 
supermultiplets and equations. For instance, Bianchi identities
from M-theory are promoted to field equations, as constraints 
on multiplets which are massless modes of M-theory bulk fields. 
With this formulation, we expect to obtain a clean, 
direct derivation of the effective supergravity suitable to 
cases more subtle than the universal modes of M-theory on $O_7$. A
first use of our formalism will be the coupling of five-brane
moduli supermultiplets in the $O_7$ case \cite{DS2}.

The organization of the paper is as follows. In Section 
\ref{secsugra}, we establish our basic supergravity 
formulation, using the bosonic bulk dynamics as starting point. The 
resulting Lagrangian is the lowest order in the 
$\kappa$-expansion. It is essentially defined by a dynamical 
Lagrangian involving tensor fields supplemented by Bianchi 
identities which are field equations of the 
theory. We discuss in detail the bosonic component expansion 
of the supermultiplet action, the question of the gravity 
frame and the generation of a superpotential. 
In Section \ref{secplanes}, we introduce the 
next order corrections: gauge multiplets and charged matter 
contributions. We show that their introduction is controlled by a 
simple modification of the four-dimensional Bianchi identities, in 
analogy with the appearance of ${\bf Z}_2$ fixed planes contributions 
in the M-theory Bianchi identities. Section 
\ref{secanom} discusses anomaly-cancelling terms. We begin by 
modifying the four-dimensional effective supergravity by adding 
terms similar to those appearing for gauge threshold corrections 
in $(2,2)$ compactifications of the heterotic string. These 
modifications can be formulated in terms of our particular 
set of multiplets directly related to M-theory bulk degrees of freedom. 
We then directly compute these anomaly-cancelling terms by 
Kaluza-Klein reduction of the 
ten-dimensional Green-Schwarz counterterms arising from
M-theory on $S^1/{\bf Z}_2$.  
Section \ref{secfinal} gives some final comments and 
conclusions, and an appendix contains our notations and
conventions. 

\section{Four-dimensional effective M-theory \\ supergravities:
bulk dynamics}
\label{secsugra}
\setcounter{equation}{0}

Our concern is compactifications of M-theory to four dimensions
preserving $N=1$ supersymmetry. Or compactifications in which
supersymmetry would break spontaneously or 
dynamically at solutions of the effective field equations. 
As a consequence,
the light (massless) modes can be described by a local
effective $N=1$ supergravity Lagrangian, to be understood in 
the sense of Wilson. 

These solutions are not as well understood as, for instance, 
Calabi-Yau compactifications of heterotic strings and it will 
prove useful to give a precise description of the aspects which 
are better known. We will then focus on a precise description 
of two aspects which are of importance in compactifications of
M-theory. Firstly, we will use a supersymmetric description of the
Bianchi identities verified by antisymmetric tensors, as arising
from M-theory or higher-dimensional supergravities. This procedure
allows to avoid multiplet ambiguities arising when duality 
transformations are performed at the bosonic level only. 
Secondly, we will use a
formulation which leaves explicitly the choice of gravity frame open.
This can be a relevant issue since an expansion, perturbative 
or not, is performed around a gravitational background which selects
a gravity frame. Standard Poincar\'e supergravity is usually written 
in the Einstein frame, in which the gravitational Lagrangian 
is $-{1\over2\kappa^2}eR$. Corrections to the lowest order 
effective action, which includes this gravitational term, induce
in general (but not always) corrections to the gravitational 
Lagrangian which affect the Einstein frame condition. 

Keeping open the choice of gravity frame suggests to use 
superconformal supergravity. And displaying Bianchi identities
explicitly in an effective Lagrangian requires using 
chiral, linear or vector supermultiplets with constraints.

The purpose of this first section is to establish our procedure
by considering the well-known `bulk dynamics', which follows
from $O_7$ compactification of eleven-dimensional 
supergravity. 

\subsection{Superconformal formalism} 

We use the superconformal formulation of $N=1$ supergravity
with a chiral compensating multiplet $S_0$ to generate 
Poincar\'e theories by gauge 
fixing\footnote{This is `old minimal' Poincar\'e supergravity 
\cite{FGKVP}.}. 
Its conformal and chiral weights are taken as $w=1$ 
and $n=1$. This formalism is particularly convenient 
to keep control of a change of frame\footnote{Mostly 
the so-called Einstein or string frames.} 
which corresponds to a different Poincar\'e gauge condition 
applied on the modulus of the scalar compensator $z_0$, 
which fixes dilatation symmetry. Up to two derivatives, a 
supergravity Lagrangian is written as\footnote{Except 
otherwise mentioned, our notation is as in ref. \cite{KU}
where also reference to the original literature can be found.}
\beq
\label{Phidef1}
{\cal L} = \left[ S_0\ov S_0 \Phi \right]_D + \left[ S_0^3 W 
\right]_F
+{1\over4}\left[f_{ab}{\cal W}^a{\cal W}^b\right]_F.
\eeq
The symbols $[\ldots]_D$ and 
$[\ldots]_F$ denote respectively the invariant $D$- and 
$F$-density {formul\ae} given by (all fermion contributions are 
omitted)
\beq
\begin{array}{rcl}
[{\cal V}]_{D} &=& e(d+{1 \over 3} cR), \crbig
[{\cal S}]_{F} &=& e(f+\ov f),
\end{array}
\eeq
where $\cal V$ is a vector multiplet with components 
$(c,\chi,m,n,b_{\mu},\lambda,d)$ and $\cal S$ a chiral 
multiplet with components $(z,\psi,f)$. The real
vector multiplet $\Phi$ with zero 
weights is a function (in the sense of tensor calculus) of the 
multiplets present in the theory, including in general the 
compensating multiplet. The holomorphic function $W$ of 
the chiral multiplets is the superpotential. 
The last term contributes to gauge kinetic terms (the chiral 
multiplet $\cal W$ is the gauge field strength for the 
gauge multiplets) and involves a holomorphic 
gauge kinetic function $f_{ab}$ of the chiral multiplets\footnote{
Gauge kinetic terms may also arise from the first term.}.
Expression (\ref{Phidef1}) provides the most general supergravity
Lagrangian up to terms with more than two derivatives and up
to terms which would contribute to kinetic terms in a fermionic
background only \cite{CFGVP1, DFKZ1}. 
Besides $S_0$ and $\cal W$, we will use chiral multiplets
with zero weights and neither $W$ nor $f_{ab}$ will depend on the
compensator. 

The chiral $U(1)$ 
symmetry of the superconformal algebra can be extended to
\beq
\label{Kahler0}
S_0 \,,\quad W \,,\quad \Phi
\qquad\longrightarrow\qquad 
\Lambda S_0 \,,\quad \Lambda^{-3}W \,,\quad
\left.(\Lambda\ov\Lambda)^{-1}\Phi 
\right|_{S_0\rightarrow\Lambda S_0}\,,
\eeq
with an arbitrary chiral multiplet $\Lambda$. This symmetry 
is at the origin of K\"ahler invariance of Poincar\'e supergravity. 
The last transformation suggests that $\log\Phi$ transforms 
as the corresponding gauge connection. Choosing  
$\Lambda= W^{1/3}$ eliminates the superpotential {\it except
if it vanishes}. One can then use a $U(1)$/K\"ahler gauge fixing in 
which the supergravity Lagrangian (\ref{Phidef1}) reads
\beq
\label{Phidef2}
{\cal L} = \left[ S_0\ov S_0 \Phi \right]_D + c \left[ S_0^3 
\right]_F +{1\over4}\left[f_{ab}{\cal W}^a{\cal W}^b\right]_F,
\eeq
with an arbitrary constant $c$ as superpotential and two 
arbitrary functions $\Phi$ and $f_{ab}$. 

The real function $\Phi$ depends on matter multiplets, which 
will be either chiral multiplets like the Calabi-Yau
universal modulus $T$, or real linear multiplets (with weights 
$w=2$, $n=0$) like the dilaton multiplet in the version 
of the theory with an antisymmetric tensor, or real vector 
multiplets ($n=0$, $w$ arbitrary) like the multiplet of gauge 
potentials ($w=0$). Vector multiplets will appear as essential 
ingredients in the effective description of $(N=1)$-preserving 
M-theory five-branes \cite{DS2}.

\subsection{Bulk Lagrangian}\label{secbulk}

The lowest order (in the $\kappa$ expansion) effective 
four-dimensional supergravity of M-theory compactified on
$O_7$ describes Kaluza-Klein 
massless modes of eleven-dimen\-sio\-nal supergravity. It
is the $S^1/{\bf Z}_2$ {\it truncation} of eleven-dimen\-sio\-nal 
supergravity on a Calabi-Yau three-fold.

In the version given by Cremmer, Julia and Scherk (CJS) 
\cite{CJS}, the Lagrangian of eleven-dimensional 
supergravity can be written as\footnote{We use the flat 
space-time metric $(-,+,+,\ldots,+)$. The gravitational constant
$\kappa_{11}$ has dimension (mass)$^{-9/2}$ and the three-index 
tensor field $C_{MNP}$ is dimensionless. Notations and 
conventions are defined in an appendix.}
\beq
\label{11dsugraCJS}
\begin{array}{rcl}
e^{-1}{\cal L}_{\rm CJS} &=& {1\over2\kappa_{11}^2}\biggl[ -R
-{1\over2\cdot4!}\,
G_{M_1M_2M_3M_4} G^{M_1M_2M_3M_4} \crbig
&&-{1\over6}\,{1\over4!4!3!}\,e^{-1}\epsilon^{M_1\ldots M_{11}}
G_{M_1M_2M_3M_4}G_{M_5M_6M_7M_8}C_{M_9M_{10}M_{11}}\biggl] \crbig
&& 
+ \,\,{\rm fermionic\,\,terms}.
\end{array}
\eeq
Omitting all fields related to the detailed geometry of 
the Calabi-Yau manifold, the particle content of the 
four-dimensional theory is the $N=1$ supergravity multiplet, 
with metric tensor $g_{\mu\nu}$, and matter multiplets 
including (on-shell) four bosons and four fermions. 
Two bosons are scalars and correspond to the dilaton and 
the `universal modulus' of the Calabi-Yau space, the 
massless volume mode. Two bosons are Kaluza-Klein modes 
of the four-form field $G$, with Bianchi identity 
$dG=0$. Explicitly, these fields and their Bianchi 
identities can be written as
\beq
\label{f4modes1}
\begin{array}{lcl}
G_{\mu\nu\rho 4}, &\qquad\qquad& 
\partial_{[\mu}G_{\nu\rho\sigma 4]} = 0,
\crbig
G_{\mu j \ov k 4} = i\, T_\mu\, \delta_{j\ov k}, && \partial_{[\mu}
T_{\nu]} = 0 . 
\end{array}
\eeq
It will prove useful to identify these fields 
with the vector components
of two real vector multiplets $V$ and $V_T$, and to impose
the Bianchi identities as field equations using the
appropriate multiplets as Lagrange multipliers. 
The bulk supergravity Lagrangian takes then the form
\beq
\label{LconfCY}
{\cal L}_{\rm B} = \left[-{1\over\sqrt2}(S_0\ov S_0V_T)^{3/2}V^{-1/2}
- (S+\ov S)V + L_TV_T\right]_D \,.
\eeq
The vector superfield $V$ (weights $w=2$, $n=0$)
includes in its components the vector field 
$v^\mu\propto \epsilon^{\mu\nu\rho\sigma}G_{\nu\rho\sigma4}$ 
for which the Bianchi identity 
is $\partial_\mu v^\mu=0$. This condition 
is a component of the (super)field equation of the chiral $S$ 
($w=n=0$) which imposes $V=L$, a real linear multiplet ($w=2$, 
$n=0$). Secondly, the vector multiplet $V_T$ ($w=n=0$) 
includes in its components $T_\mu$. The supersymmetric extension
of its Bianchi identity $\partial_{[\mu}T_{\nu]} = 0$ 
is enforced by the real linear multiplet $L_T$,
which implies $V_T=T+\ov T$, with a chiral weightless 
multiplet $T$. The usefulness of obtaining Bianchi 
identities via field equations will become apparent 
with the introduction of higher orders in the $\kappa$ 
expansion. At this stage of the discussion however, it 
gives a formulation of the familiar duality relating 
scalars and antisymmetric tensors or, for superfields, chiral 
and linear multiplets. 

Solving in Eq. (\ref{LconfCY}) for the Lagrange multipliers 
$S$ and $L_T$ leads to the `standard form' of the bulk 
four-dimensional Lagrangian
\cite{CFV, DQQ}
\beq
\label{4dLCJS}
{\cal L}_{\rm B,l} = -{1\over\sqrt2}\left[ 
\left(S_0\ov S_0 \, e^{-\hat K/3}\right)^{3/2}
L^{-1/2}\right]_D \,,
\eeq
or, as defined in Eq. (\ref{Phidef1}), 
\beq
\label{phiis1}
\Phi = -\left({2L\over S_0\ov S_0 }\right)^{-1/2}e^{-\hat K/2}.
\eeq
The K\"ahler potential for the volume modulus $T$ is
\beq
\label{hatKis}
\hat K  = -3\log(T + \ov T).
\eeq
We will see again below that this standard form is 
naturally obtained by direct reduction of the CJS version 
of eleven-dimensional supergravity on $O_7$.
Clearly, theory (\ref{4dLCJS}) is also the Calabi-Yau truncation 
of ten-dimensional $N=1$ pure supergravity \cite{CFV}.
Notice that $\hat K$ can be regarded as the K\"ahler connection
for symmetry (\ref{Kahler0}), with transformation
\beq
\label{Kahler1}
\hat K \qquad\longrightarrow\qquad
\hat K +3\log \Lambda +3\log\ov\Lambda,
\eeq
such that $S_0\ov S_0\, e^{-\hat K/3}$ is chiral/K\"ahler 
invariant.

A theory with a linear multiplet is in principle  
dual to an equivalent Lagrangian with the linear multiplet 
replaced by a chiral one. In our case, solving for $V$ 
and $L_T$ in expression (\ref{LconfCY}) leads to
\beq
\label{bulkchiral}
\begin{array}{rcl}
{\cal L}_{\rm B,c} &=& -{3\over2}\left[S_0\ov S_0\, e^{-K/3}
\right]_D, \crbig
K &=&  -\log(S+\ov S)+\hat K \,\,=\,\,-\log(S+\ov S)
-3\log(T+\ov T).
\end{array}
\eeq
This familiar chiral form \cite{W4dsugra} is not the 
most useful as long as one insists on the four-dimensional 
translation of eleven-dimensional Bianchi identities.

Notice that one can obtain another equivalent form of the 
Lagrangian (\ref{LconfCY}) by choosing to solve for $S$ 
and $V_T$. In this case, the Calabi-Yau modulus is 
described by a linear multiplet $L_T$. This form will not 
be useful since it is known that one-loop string
corrections in general break the chiral-linear duality for this
modulus: they involve holomorphic functions of $T$ in a $F$-density 
which are intrinsically chiral \cite{DKL}. 
Finally, there is an obstruction when trying to solve 
for $V$ and $V_T$ and one cannot write an expression 
in terms of the chiral $S$ and the linear $L_T$. 

Before turning to explicit component expressions, we 
should discuss the choice of Poincar\'e frame, and introduce 
the expansion in the four-dimensional gravity coupling 
$\kappa$, which effectively corresponds to the low-energy 
expansion of M-theory in powers of the eleven-dimensional 
gravitational constant $\kappa_{11}$. 

\subsubsection{Einstein frame}\label{secEinstein}

To gauge-fix dilatations, we impose as usual a condition on the
Einstein term appearing in the superconformal supergravity 
Lagrangian. According to the component expression for the 
$D$-density and the tensor calculus of superconformal 
multiplets \cite{KU}, the Einstein term included in 
$[S_0\ov S_0\Phi]_D$ is \cite{FGKVP, DQQ}
\beq
\label{Eins1}
{\cal L}_{\rm E} = -{1\over2}eR\left[ -{2\over3}z_0\ov z_0\left(\Phi - 
{1\over2}\sum_i w_iC_i{\partial\Phi\over\partial C_i}\right) 
\right],
\eeq
where $w_i$ is a Weyl weight, the sum is taken over the 
linear ($w_i=2$) and vector ($w_i$ arbitrary) multiplets, 
and $z_0$, $\Phi$ and $C_i$ are the lowest, scalar 
components of respectively $S_0$, $\Phi$ and of the vector 
or linear multiplets\footnote{We use in general the same 
notation for the lowest component of $\Phi$ and the multiplet 
itself. Also, in expression (\ref{Eins1}), we use the component 
expansion of vector multiplets indicated below [Eq.(\ref{comp1})], 
which differs in its highest component from ref. \cite{KU}.}.

Applied to the bulk Lagrangian (\ref{LconfCY}), 
expression (\ref{Eins1}) leads to
\beq
\label{Eins2}
{\cal L}_{\rm E} = -{1\over2}eR\left[ (z_0\ov z_0 C_T)^{3/2}(2C)^{-1/2} 
\right] .
\eeq
As they should, the terms introduced to impose Bianchi 
identities do not contribute. The Einstein frame is then 
selected by the dilatation gauge condition 
\beq
\label{Eins3}
\kappa^{-2} = (z_0\ov z_0 C_T)^{3/2}(2C)^{-1/2}.
\eeq
It will be convenient to introduce the (composite) real vector 
multiplet
\beq
\label{Upsilonis}
\Upsilon = (S_0\ov S_0 V_T)^{3/2} (2V)^{-1/2},
\eeq
with conformal weight two. In the Poincar\'e theory and in 
the Einstein frame, its lowest component is precisely equal 
to $\kappa^{-2}$. With this definition, the bulk Lagrangian 
becomes simply
\beq
\label{Lbulkupsilon}
{\cal L}_{\rm B} = \left[ -\Upsilon - (S+\ov S)V
+ L_TV_T\right]_D,
\eeq
and the equation of motion for $V$ (the chiral-linear 
duality equation)
\beq
\label{Veom}
2V(S+\ov S) = \Upsilon
\eeq
indicates that the Einstein Lagrangian also reads 
$$
{\cal L}_{\rm E} = -(2C\Re s) \, eR.
$$
In the Einstein frame (Planck units), $\Re s = (
4\kappa^2 C)^{-1}$.

Eq. (\ref{Veom}) is compatible with the standard relation
of heterotic strings\footnote{
We use $\langle\ldots\rangle$ for a background value.} 
$2\kappa^2 \langle\Re s\rangle = \alpha^\prime$
if one identifies $2\langle C\rangle = 1/\alpha^\prime$.
This equation defines string units, in which
\beq
\label{str2}
{\cal L}_{\rm E} = - {e^{-2\varphi}\over\alpha^\prime}\, eR,
\eeq
with a dilaton given by $e^{-2\varphi} = \Re s$.

\subsubsection{Modified Bianchi identities and $\kappa$-expansion}
\label{seckappaexp}

Compactification of M-theory on $S^1/{\bf Z}_2$ is commonly 
discussed in an expansion in powers of $\kappa_{11}$. 
Compactification on $O_7$ can similarly be formulated with 
$\kappa$ as expansion parameter. In the upstairs version, 
Bianchi identities are modified at the ten-dimensional planes 
fixed by $S^1/{\bf Z}_2$. Suppose now that we modify the 
four-dimensional supersymmetric Bianchi identities of the 
bulk Lagrangian in the following way:
\beq
\label{modif1}
{\cal L}_{\rm B} \quad\longrightarrow \quad 
{\cal L} = \biggl[ -\Upsilon 
- (S+\ov S)(V + \Delta_V)
+ L_T(V_T +\Delta_T) \biggr]_D\,,
\eeq
with two composite vector multiplets $\Delta_V$ ($w=2$, $n=0$) 
and $\Delta_T$ ($w=n=0$). Solving for the Lagrange 
multipliers now leads to
$$
V= L-\Delta_V, \qquad
V_T=T+\ov T- \Delta_T.
$$
The Lagrangian to first order in these modifications reads then
\beq
\label{modif2}
\begin{array}{rcl}
{\cal L} &=& {\cal L}_{\rm B} - \left[{\Upsilon\over 2V} \Delta_V 
-{3\over2}{\Upsilon\over V_T}
\Delta_T\right]_D 
\crbig
&=& 
{\cal L}_{\rm B} - \left[(S+\ov S)\Delta_V 
-{3\over2V_T}(\Upsilon\Delta_T)\right]_D,
\end{array}
\eeq
with $V$ and $V_T$ respectively replaced by $L$ 
and $T+\ov T$ in these expressions.
The multiplets $\Delta_V$ and $\Upsilon\Delta_T$, 
with `canonical' dimension $w=2$, appear at order 
$\Upsilon^0 \sim \kappa^0$, in comparison with bulk terms 
of order $\Upsilon\sim\kappa^{-2}$. This is the relation 
with the expansion in powers of $\kappa_{11}$ of M-theory 
in the low-energy limit. In M-theory compactification,
the multiplets $\Delta_V$ and $\Delta_T$ can thus be obtained
either by considering the modified Bianchi identities on $O_7$, 
formulated as in Eq. (\ref{modif1}), or from corrections to the 
Lagrangian of eleven-dimensional supergravity on $O_7$, as in
expression (\ref{modif2}).

\subsubsection{Component expressions}

To analyze the Lagrangian (\ref{LconfCY}), 
we will need to define some notations. Since we will only 
explicitly consider the bosonic sector of the theory, all 
fermions in the $N=1$ supermultiplets will be omitted. 
Since also we are concerned with Poincar\'e supergravity, 
we will immediately gauge-fix the superconformal
symmetries not contained in $N=1$ Poincar\'e supersymmetry, 
with one exception, dilatation symmetry: we want to keep the 
freedom of a frame choice as explicit as possible. These 
assumptions imply in particular that superconformal covariant 
derivatives reduce in general to
$D^c_\mu\phi = D_\mu\phi - {i\over2}nA_\mu\phi$ for a complex field 
with chiral weight $n$ and to
\beq\label{Boxis}
\Box^c\varphi = \Box\varphi + {1\over6}w\varphi\,R,
\eeq
if $w$ is the Weyl weight and $\varphi$ real. 
The gauge boson $A_\mu$ of chiral 
$U(1)$ symmetry is auxiliary and $D_\mu$ and $\Box$ would be 
covariantized with respect to Poincar\'e symmetries. 

We will use the following components for the various 
superconformal multiplets appearing in Lagrangian (\ref{LconfCY}):
\beq
\label{comp1}
\begin{array}{rcl}
V &=& (C,0,H,K,v_\mu,0, d-\Box C-{1\over3}CR), 
\crbig 
V_T &=& (C_T,0,H_T,K_T,T_\mu,0, d_T-\Box C_T), \crbig 
S &=& (s,0,-f,if,i\partial_\mu s,0, 0), \crbig 
L_T &=& (\ell_T,0,0,0,t_\mu,0, -\Box\ell_T-{1\over3}\ell_T R), 
\qquad\quad 
t_\mu={e\over2}\epsilon_{\mu\nu\rho\sigma}\partial^\nu t^{\rho\sigma},
\crbig
S_0 &=& (z_0,0,-f_0,if_0,iD^c_\mu z_0,0, 0). 
\end{array}
\eeq
The role of the Lagrange multipliers $S$ and $L_T$ follows from
\beq
\label{SLT}
\begin{array}{rcl}
e^{-1}[(S+\ov S)V]_D &=& 2(\partial^\mu\Im s)v_\mu -2 
\partial^\mu(\Re s\,
\partial_\mu C) + 2d\Re s \crbig
&& -f(H-iK) -\ov f(H+iK)
\crbig
&=& -2\Im s \partial^\mu v_\mu + 2d\Re s
-f(H-iK) -\ov f(H+iK) \crbig
&& +~{\rm derivative}, \crbig
e^{-1}[L_TV_T]_D &=& 
\ell_T(d_T-\Box C_T) 
- {e\over2}\epsilon_{\mu\nu\rho\sigma}(\partial^\mu T^\nu) 
t^{\rho\sigma} + {\rm derivative}.
\end{array}
\eeq
Solving for the components of $S$ leads to $\partial^\mu v_\mu=d=H=K=0$,
and $V$ is a linear multiplet. Solving for the components 
of $L_T$ leads to $d_T-\Box C_T= \partial_{[\mu} T_{\nu]}=0$, 
and $V_T$ can be written as $T+\ov T$, with a chiral 
multiplet $T$ (zero weights)\footnote{
The components are: $C_T=2\Re T$, $T_\mu=-2\partial_\mu\Im T$, 
$H_T=-2\Re f_T$, $K_T=-2\Im f_T$.}.

Since one can always write $v_\mu 
={e\over6}\epsilon_{\mu\nu\rho\sigma}
v^{\nu\rho\sigma}$, we have generated with $\Im s$ and $t_{\mu\nu}$
the Bianchi identities
$$
\partial_{[\mu}v_{\nu\rho\sigma]} 
= \partial_{[\mu} T_{\nu]} = 0\,.
$$
A modification of these Bianchi 
identities, as induced by $S^1/{\bf Z}_2$ compactification or by 
five-brane couplings will then be phrased as a modification 
of the supermultiplets appearing multiplied by $S+\ov S$ or 
$L_T$ in Eqs. (\ref{SLT}). 

To complete the identification of the four-dimensional 
supergravity (\ref{LconfCY}) with the modes of 
eleven-dimensional supergravity we need its complete bosonic 
expansion, which after solving for the
chiral $U(1)$ auxiliary field $A_{\mu}$ reads:
\beq
\label{bulkcomp}
\begin{array}{rcl}
e^{-1}{\cal L}_{\rm B} &=&
   - {1 \over 2} \Upsilon R 
   + {1 \over 4} \Upsilon C^{-2} v_\mu v^\mu 
   - {3 \over 4} \Upsilon C_T^{-2} T_\mu T^\mu 
   - t_\mu T^\mu + 2 \Im s \partial_\mu v^\mu \crbig
&& - {1 \over 4} \Upsilon C^{-2} 
	(\partial_\mu C)(\partial^\mu C)  
   - {3 \over 4} \Upsilon C_T^{-2} 
	(\partial_\mu C_T)(\partial^\mu C_T)
   \crbig
&& + d({1 \over 2} \Upsilon C^{-1} - 2 \Re s) 
   + (d_T -  \Box C_T) (\ell_T - {3 \over 2} \Upsilon C_T^{-1}) 
\crbig 
&& + {1 \over 2} (\partial_\mu \Upsilon) 
	[\partial^\mu \log C + \partial^\mu \log \Upsilon ] \crbig
&& + e^{-1}{\cal L}_{\rm AUX.} + \rm{derivative},
\end{array}
\eeq
where $\Upsilon = (z_{0} \ov z_{0}C_T)^{3/2} (2C)^{-1/2}$, and
\beq
\label{bulkauxcomp}
\begin{array}{rcl}
e^{-1}{\cal L}_{\rm AUX.} &=& 
- {1 \over 4} \Upsilon C^{-2} (H+iK) (H-iK) 
+ {3 \over 4} \Upsilon C_T^{-2} (H_T+iK_T) (H_T-iK_T) \crbig
&& + f(H-iK) + \ov f(H+iK).
\end{array}
\eeq
The last equality is obtained after solving for the
$f_{0}$ component of $S_{0}$ \footnote{Note that the result 
would be different with a superpotential.}.

The above component expansion of the bosonic Lagrangian is useful
because it explicitly displays the dependence on the gauge choice 
for dilatation symmetry. The gravitational constant is the 
field-dependent quantity $\Upsilon$. Choosing a Poincar\'e 
frame amounts to impose the value of this quantity,
and to use this condition to eliminate $z_0$. Notice also 
that the choice of the phase of $z_0$ is a gauge condition 
for the chiral internal $U(1)$ superconformal symmetry. 
With the exception of $z_0$, all bosons are $U(1)$-neutral. 
As a consequence, the bosonic Lagrangian only depends on
the modulus of $z_0$.

Taking the Einstein frame, $\Upsilon = \kappa^{-2}$, and 
solving for the components of $S$ and $L_T$ leads to
\beq
\label{bulkcompfinal}
\begin{array}{rcl}
e^{-1}{\cal L}_{\rm B} &=&
   - {1 \over {2\kappa^2}}  R 
- {1 \over {4\kappa^2}} C^{-2} 
	[ (\partial_\mu C)(\partial^\mu C) - v_\mu v^\mu ] \crbig
&& - {3 \over {4\kappa^2}} C_T^{-2} 
	[ (\partial_\mu C_T)(\partial^\mu C_T) + T_\mu T^\mu ],
\end{array}
\eeq
with $v_\mu={e\over2}\epsilon_{\mu\nu\rho\sigma}
\partial^{\nu}b^{\rho\sigma}$ since $V$ is a linear multiplet,
$C_T=2\Re T$ and $T_\mu=-2\partial_\mu\Im T$ since $V_T=T+\ov T$.
This Lagrangian is to be compared with the reduction of 
the CJS version of eleven-dimensional supergravity 
(\ref{11dsugraCJS}). 
The ${\bf Z}_2$ orbifold projection eliminates all
states which are odd under $x^4\rightarrow -x^4$, and since we 
disregard massless modes related to the detailed Calabi-Yau 
geometry, the reduction of the $D=11$ space-time metric is
\beq
\label{D11metric}
g_{MN} = \pmatrix{
e^{-\gamma} e^{-2\sigma} g_{\mu\nu} & 0 & 0 \cr 
0 & e^{2\gamma} e^{-2\sigma} & 0 \cr
0 & 0 & e^{\sigma} \delta_{i\ov j} \cr}.
\eeq
The $SU(3)$--invariant tensor $\delta_{i\ov j}$ refers to 
complex coordinates on the Calabi-Yau space. 
The surviving components of the four-index tensor $G_{MNPQ}$ 
are only $G_{\mu\nu\rho4}$ and $G_{\mu i\ov j4}$, with
$$
G_{\mu\nu\rho4} = 3\partial_{[\mu}C_{\nu\rho]4}, 
\qquad G_{\mu i\ov j4}=\partial_\mu C_{i\ov j4}, \qquad 
C_{i\ov j4}= i a(x)\,\delta_{i\ov j},
$$
and the four-dimensional Lagrangian for these fields reads
\beq
\label{CJSbulkcomp}
\begin{array}{rcl}
e^{-1}{\cal L}_{\rm CJS} &=& -{1\over2\kappa^2}R
-{1\over4\kappa^2}\Bigl[9(\partial_\mu\sigma)
(\partial^\mu\sigma)+{1\over6}e^{6\sigma} G_{\mu\nu\rho4}G^{\mu\nu\rho4}\Bigr]
\crbig
&&
-{3\over4\kappa^2}\Bigl[(\partial_\mu\gamma)(\partial^\mu\gamma)
+e^{-2\gamma} (\partial_\mu a)(\partial^\mu a) \Bigr].
\end{array}
\eeq
In this expression, $\kappa$ is the four-dimensional 
gravitational coupling 
$$
\kappa^2 = {\kappa_{11}^2\over V_7},
$$
$V_7=V_1V_6$ being the volume of the compact space 
$S^1 \times X_6$.

At this stage, the identification of the bosonic components 
$C$ and $b_{\mu\nu}$ of $V$, $C_T$ and $T_\mu$ of $V_T$ with the
bulk fields $\sigma$, $\gamma$, $C_{\mu\nu4}$ and $a$ can 
only be determined up to a proportionality constant for each 
multiplet. We will define these constants later on 
from the couplings of $C$ and $C_T=2\Re T$ 
to charged matter and gauge fields, to obtain:
\beq
\label{CJScompident}
\begin{array}{ll}
4\kappa^2 C = {\lambda^2\over V_6}e^{-3\sigma}, \qquad 
&4\kappa^2 b_{\mu\nu} = {\lambda^2\over V_6}C_{\mu\nu 4}, 
\crbig
C_T = 2{\lambda^2\over V_6}e^\gamma \,\,=\,\, 2 \Re T, 
\qquad &T_\mu = -2{\lambda^2\over V_6} \partial_\mu a
\,\,=\,\, - 2 \partial_\mu \Im T.
\end{array}
\eeq
The quantity $\lambda$ is the gauge coupling constant on the
${\bf Z}_2$ fixed planes and $\lambda^2/V_6$ is a dimensionless
number which will actually never appear in the four-dimensional
effective theory. 

\subsubsection{Bianchi identities and symmetries}

In the bulk Lagrangian (\ref{LconfCY}), the terms 
$[-(S+\ov S)V + L_TV_T]_D$ impose in particular the Bianchi 
identities (\ref{f4modes1}). They are certainly 
invariant under 
$$
\begin{array}{rcll}
V&\quad\longrightarrow\quad& V+L , \qquad & L\,\,
{\rm linear},\crbig V_T&\quad\longrightarrow\quad& 
V_T+ T+\ov T , \qquad & T\,\,{\rm chiral}.
\end{array}
$$
These symmetries are the supersymmetric extensions of the 
gauge invariances of  Bian\-chi identities, 
$\delta G_{\mu\nu\rho4}=3\partial_{[\mu}\Lambda_{\nu\rho]}$
and $\delta G_{\mu i\ov j4} = i\partial_\mu\Lambda\,\delta_{i\ov j}$. 
Solving for $S$ and $L_T$ implies then that $V$ and 
$V_T$ are `pure gauge', $V=L$ and $V_T=T+\ov T$. The last 
equation defines $V_T$ up to a holomorphic redefinition 
of $T$, $T\rightarrow f(T)$. This redefinition is a 
symmetry of the bulk Lagrangian if the function $\Phi$ 
simultaneously transforms as in (\ref{Kahler0}), with 
$\Lambda$ a holomorphic function of $T$. 
The equation for the invariance of $S_0\ov S_0 V_T$ is
$$
f(T)+\ov f(\ov T) = {T+\ov T\over\Lambda(T)\ov \Lambda(\ov T)}\,,
$$
and its solution is clearly $Sl(2,\R)$ symmetry, 
\beq
\label{modinv1}
T \qquad\longrightarrow\qquad f(T)= {aT-ib\over icT+d},
\qquad\qquad ad-bc=1,
\eeq
the modular invariance of $T$ (T-duality), extended to a 
continuous symmetry at the lowest order. 

This chiral symmetry is generically anomalous: in the presence
of a $N=1$ super-Yang-Mills sector, with or without chiral matter, 
mixed anomalies arise in the triangle diagram for two 
gauge bosons and one connection $-3\log(T+\ov T)$ for 
$Sl(2,\R)$ symmetry. This anomaly is cancelled in particular 
by a Green-Schwarz mechanism as was demonstrated in the 
effective Lagrangian description of gauge thresholds 
\cite{DFKZ1} calculated at one-loop 
for $(2,2)$ compactifications of the heterotic strings 
\cite{DKL}. We will see below that this phenomenon is also 
a useful tool in the construction of effective supergravities 
of M-theory compactifications.  

\subsubsection{Superpotential}\label{secsuperpot}

The standard reduction of eleven-dimensional supergravity with 
unbroken $N=1$ supersymmetry does not 
generate a superpotential. This fact is however not a direct 
consequence of the eleven-dimensional Bianchi identity 
or of the Calabi-Yau and $S^1/{\bf Z}_2$ symmetries. In principle,
the Bianchi identity $\partial_{[M}G_{NPQR]}=0$ allows a solution
\beq
\label{Fijk4}
G_{ijk4} = 2i\kappa^{-1}h \epsilon_{ijk}, 
\qquad\qquad
G_{\ov {ijk}4} = -2i\kappa^{-1}h \epsilon_{\ov{ijk}}.
\eeq
In these equations, $h$ is a constant chosen real and 
$\epsilon_{ijk}$ is the $SU(3)$--invariant Calabi-Yau tensor. 
The Lagrangian term $-{e\over2\kappa_{11}^2} {1\over48}
G_{MNPQ}G^{MNPQ}$ leads then to a contribution
$$
-{e\over\kappa^4} C_T^{-3}(2\kappa^2C) h^2
$$
in the four-dimensional effective supergravity. This contribution 
corresponds to the addition of a superpotential term
$$
[ihS_0^3]_F 
$$
to the bulk Lagrangian, a contribution which however breaks 
supersymmetry \cite{DIN1}. Since we have insisted in writing 
Lagrangians in which all Bianchi identities are field 
equations, we prefer instead to use  
\beq
\label{superpot1}
[U(W+\ov W)]_D + [S_0^3\, W]_F.
\eeq
The field equation of the vector multiplet $U$ (weights $w=2$, 
$n=0$) implies that the chiral multiplet $W$ ($w=n=0$)
is an arbitrary imaginary constant, which can be zero
and supersymmetry stays unbroken, or non-zero.

With the addition (\ref{superpot1}) of a superpotential, 
the bulk Lagrangian takes its final `off-shell' form
\beq
\label{LcompCYfinal}
{\cal L}_{\rm B} = \left[ -\Upsilon
-(S+\ov S)V + L_TV_T + U(W+\ov W) \right]_D
+ [S_0^3 W]_F,
\eeq
in which the Bianchi identities of eleven-dimensional 
supergravity are translated into field equations of the 
Lagrange multipliers $S$, $L_T$ and $U$. At this stage, 
the introduction of these multiplets is not 
fascinating. This approach encodes simply the 
(Poincar\'e) dualities relating antisymmetric tensors (in 
linear multiplets) and scalars (in chiral 
multiplets), and the Bianchi identity for the superpotential is 
trivial. But this procedure will prove useful and 
informative in the forthcoming sections.

\section{Fixed planes: gauge and matter contributions}\label{secplanes}
\setcounter{equation}{0}

In this section, we start with the well-known effective $N=1$
four-dimensional supergravity for symmetric $(2,2)$ 
compactifications of heterotic strings. We then rewrite 
this theory in a form where explicit Bianchi identities allow 
a direct comparison with $O_7$ compactification of M-theory, 
in the so-called upstairs formulation \cite{HW1, HW2}.

The dependence on charged matter (in chiral multiplets collectively 
denoted by $M$, with $w=n=0$) and gauge multiplets (vector multiplet 
$A$, in the adjoint representation, with $w=n=0$) of
the effective supergravity theory for Calabi-Yau
compactifications of heterotic strings is well-known \cite{W4dsugra,
DIN, BFQ}, at least for the `universal' matter multiplets arising 
from the simplest Calabi-Yau modes of the ten-dimensional 
super-Yang-Mills fields. Information on the non-trivial 
harmonic modes is more subtle \cite{DKL2}, as for generic 
Calabi-Yau moduli. In the chiral formulation, Eq. 
(\ref{bulkchiral}) becomes
\beq
\label{matter1}
\begin{array}{rcl}
{\cal L}_{\rm c} &=& -{3\over2}\left[S_0\ov S_0 e^{-K/3}\right]_D
+ [{1\over4}S{\cal WW} + S_0^3W]_F, \crbig
K &=& -\log(S+\ov S)-3\log(T+\ov T - 2\ov Me^AM), \crbig
W &=& \alpha M^3.
\end{array}
\eeq
For notational simplicity, we omit traces over the gauge group 
representation and their normalisation factors. 
The chiral multiplet ${\cal W}$ is the gauge field-strength 
for $A$ ($w=n=3/2$). Since the gauge group is in general not 
simple, 
\beq
\label{Wdef}
{\cal WW} = \sum_a c^a{\cal W}^a{\cal W}^a,
\eeq
with a (real) coefficient $c^a$ for each simple or abelian 
factor\footnote{Corresponding to Kac-Moody levels
in superstrings. All coefficients can be equal to one, as 
with the `standard embedding', but our discussion is not 
affected by their presence.}. The superpotential should be 
understood as a gauge invariant trilinear interaction with 
coupling constant $\alpha$ defined as an integral over the 
Calabi-Yau space.
In the linear multiplet version, the equivalent expression is
\cite{CFV, DQQ}
\beq
\label{matter2}
{\cal L}_{\rm l} = -{1\over\sqrt2}\left[ (S_0\ov S_0)^{3/2}
\hat L^{-1/2} e^{-\hat K/2}\right]_D +[\alpha S_0^3 M^3]_F \,.
\eeq
With respect to Eq. (\ref{4dLCJS}), gauge and matter dependence 
arises in modifications of the linear multiplet $L$ (to $\hat L$) 
and of $\hat K$: the new modulus and matter K\"ahler potential is
\beq
\label{hatKis2}
\hat K = -3\log(T+\ov T - 2\ov M e^A M)  \,,
\eeq
instead of Eq. (\ref{hatKis}) and
\beq
\label{Lhatis}
\hat L = L - 2\Omega,
\eeq
where $\Omega(A)$ is the Chern-Simons vector multiplet ($w=2$, 
$n=0$), defined by \footnote{
In global Poincar\'e supersymmetry, $\Sigma(\Omega) = -{1\over4}
\ov{DD} \Omega$. A linear multiplet is defined by the condition
$\Sigma(L)=0$.} 
\beq
\label{CSdef}
\Omega = \sum_a c^a\Omega^a, \qquad\qquad
\Sigma(\Omega^a) = {1\over16}{\cal W}^a{\cal W}^a.
\eeq

Insisting as before on Bianchi identities, both forms 
(\ref{matter1}) and (\ref{matter2}) are equivalent to 
\beq
\label{matter3}
\begin{array}{rcl}
{\cal L} &=& \biggl[ -(S_0\ov S_0V_T)^{3/2}(2V)^{-1/2}
-(S+\ov S)(V+2\Omega) \crbig
&& + L_T(V_T+2\ov Me^AM) + \{U(W-\alpha M^3)+ {\rm c.c.}\} 
\biggr]_D + [S_0^3 W]_F \crbig
&=& \biggl[ -\Upsilon
-(S+\ov S)(V+2\Omega) + L_T(V_T+2\ov Me^AM) \biggr]_D
+ [S_0^3(ih +\alpha M^3)]_F.
\end{array}
\eeq
Supersymmetric vacua have $h=0$. 
As before, solving in the last expression for $S$ and 
$L_T$ imposes respectively $V=L-2\Omega=\hat L$ and 
$V_T=T+\ov T-2\ov Me^AM$, leading to Eq. (\ref{matter2}). 
Alternatively, solving for $V$ and $L_T$ leads back to 
the chiral form (\ref{matter1}), with the tensor 
calculus identity
\beq
\label{DFident}
-2[(S+\ov S)\Omega]_D = {1\over4}\sum_ac^a[S{\cal W}^a
{\cal W}^a]_F + {\rm derivative},
\eeq
which follows from Eq. (\ref{CSdef}) and the definition 
of the $F$-density, $[\Sigma(\ldots)]_F = -[\ldots]_D$.

This reformulation of the gauge invariant Lagrangian suggests 
some remarks. 
First\-ly, it enhances the importance of Chern-Simons multiplets
in superstring effective actions: gauge fields {\it and} matter
fields couple to the bulk Lagrangian using a Chern-Simons
multiplet.
The gauge Chern-Simons multiplet $\Omega$ is defined by Eq. 
(\ref{CSdef}), which indicates that its chiral projection 
$\Sigma(\Omega)$ is the chiral multiplet for the kinetic 
super-Yang-Mills Lagrangian. Similarly for chiral matter, 
the kinetic Wess-Zumino Lagrangian can be written as 
$[S_0\ov S_0\,\ov Me^AM]_D = 
-[\Sigma(S_0\ov S_0\,\ov Me^A M)]_F$, defining 
\beq
\label{matterCSis}
\Omega_M = S_0\ov S_0 \ov Me^A M
\eeq
as a matter Chern-Simons multiplet ($w=2$, $n=0$) which
then couples to $S_0\ov S_0 V_T$ as $\Omega$ couples to $V$.

Secondly, the Chern-Simons vector multiplet $\Omega(A)$ is 
not gauge invariant: its variation is a linear multiplet. 
Then, the variation of $[(S+\ov S)\Omega]_D$ is a derivative 
and $V$ remains gauge invariant. When solving for $S$, it 
simply follows that $\hat L$ is gauge invariant and that the 
linear multiplet transforms as
\beq
\label{Lvaria}
\delta L = 2\delta\Omega.
\eeq

Finally, expression (\ref{matter3}) shows that all gauge 
and chiral matter contributions can be viewed as the 
supersymmetrization of modified Bianchi identities
imposed by $S$, $L_T$ and $U$. This is equally true in the 
ten-dimensional supergravity--Yang-Mills system: the
curl of the antisymmetric tensor field is modified by 
Chern-Simons contributions which are supersymmetry partners 
of the super-Yang-Mills Lagrangian \cite{CM}. 
This observation provides the link to the approach 
based on M-theory on $O_7$, in which the ${\bf Z}_2$--fixed planes
carrying the Yang-Mills fields induce because of supersymmetry
modifications to the Bianchi identity of the four-form field 
strength of eleven-dimensional supergravity. 

In the effective supergravity of M-theory on $O_7$ (`upstairs 
formulation'), the various components of the Lagrangian 
(\ref{matter3}),
$$
\begin{array}{rcl}
{\cal L} &=& \biggl[ -(S_0\ov S_0V_T)^{3/2}(2V)^{-1/2}
-(S+\ov S)(V+2\Omega)  \crbig
&& + L_T(V_T+2\ov Me^AM) + \{ U(W-\alpha M^3)+ {\rm c.c.}\} 
\biggr]_D + [S_0^3 W]_F ,
\end{array}
$$
have the following origin. As already discussed at length, the first
term is the bulk supergravity contribution. Then $[(S+\ov S)
(V+2\Omega)]_D$ is the supersymmetrization of the Bianchi identity
verified by the component $G_{\mu\nu\rho4}$ of the four-form field, 
modified by gauge contributions on the fixed planes. Similarly,
$[L_T(V_T+2\ov Me^AM)]_D$ and $[U(W-\alpha M^3)+{\rm c.c.}]_D$
are respectively the supersymmetric extensions of the 
Bianchi identities of $G_{\mu j\ov k 4}$ and $G_{ijk4}$, 
when fixed plane contributions are included. Thus, all fixed plane 
contributions are given at this order by the supersymmetrization
of Bianchi identities, as obtained by direct $O_7$
truncation of the eleven-dimen\-sio\-nal identities 
\cite{HW1, HW2}. 

At this point, the gauge coupling constant for each simple or 
abelian factor $a$ in the gauge group appears to be
\beq
\label{gis1}
{1\over g_a^2} = c^a \Re s = {c^a\Upsilon\over 4C},
\eeq
$s$ and $C$ being respectively the lowest scalar component of 
the chiral $S$ and the vector $V$ (or the linear $L$). At this
order, $g_a$ is the tree-level wilsonnian and
physical\footnote{The coefficient of 
$-{1\over4}F_{\mu\nu}^aF^{a\mu\nu}$
in the generating functional of one-particle irreducible Green's 
functions.} gauge coupling. The
second equality is the lowest component of 
the equation of motion of the vector multiplet 
$V$, Eq. (\ref{Veom}). In the Einstein frame, 
$\Re s= (4\kappa^2 C)^{-1}$.

It is clear, as already observed \cite{ref1}--\cite{LOW1},
that as far as the structure of the four-dimensional
effective supergravity is concerned, the same information follows
from $O_7$ compactification of M-theory at the next to lowest 
order in the $\kappa$ expansion and from Calabi-Yau 
compactifications of the heterotic strings, at zero string
loop order. 

Notice that Eq. (\ref{gis1}) defines $c^a\Re s$ as the coefficient
of gauge kinetic terms. It defines then this field 
in terms of the gauge kinetic action on the ten-dimensional 
${\bf Z}_2$ fixed planes,
\beq 
\label{Sgauge}
{\cal S}_{\rm gauge} =
-{1\over4\lambda^2}\int_{M_{10}}d^{10}x\, e_{10}
{\rm tr}\, F_{AB}F^{AB},
\eeq
reduced on $X_6$. This action is also at the origin of 
charged matter kinetic terms, which in the effective 
supergravity read
$$
-{e\over\kappa^2}\,{\partial^2\hat K\over\partial\ov M\partial M}
\,(D_\mu \ov M)(D^\mu M) =
-{6e\over\kappa^2C_T}(D_\mu \ov M)(D^\mu M) + \ldots
$$
The Calabi-Yau reduction of the action 
(\ref{Sgauge}) provides then the identification of $C$ and 
$C_T$ in terms of the bulk fields $\sigma$ and $\gamma$ 
appearing in the metric tensor (\ref{D11metric}). These 
results have already been displayed in Eq. (\ref{CJScompident}).

\section{Anomaly-cancelling terms}\label{secanom}
\setcounter{equation}{0}

In the ten-dimensional
heterotic string, cancellation of gauge and gravitational 
anomalies is a one-loop effect in string 
or effective supergravity perturbation theory. 
In the low-energy effective action description, we should then
distinguish the Wilson effective supergravity 
from the standard effective action ${\cal S}_\Gamma$, 
defined as the generating functional of one-particle 
irreducible Green's functions. The latter action can 
be obtained in a diagrammatic expansion built from the 
Wilson Lagrangian ${\cal L}$, itself obtained from string 
perturbation theory as an expansion
$$
{\cal L} = {\cal L}^{(0)} + {\cal L}^{(1)} +\ldots,
$$
the subscript being the string-loop order. 
The expressions given in the previous sections were for 
${\cal L}^{(0)}$, or for the tree-level ${\cal S}_\Gamma$. 
At the string one-loop level, ${\cal S}_\Gamma$
includes tree and one-loop diagrams generated by the Feynman 
rules of the tree-level Wilson Lagrangian ${\cal L}^{(0)}$.
These include anomalous loop diagrams. 
It also includes tree diagrams generated by the
Green-Schwarz counterterm \cite{GS} introduced in
${\cal L}^{(1)}$ to cancel the anomalies generated by ${\cal L}^{(0)}$.
The mechanism for symmetry restoration implies that 
${\cal L}^{(1)}$ is not invariant under the restored symmetry.

In four space-time dimensions, the nature of the cancelled 
anomalies is known from studies of $(2,2)$ compactifications 
of heterotic strings in the Yang-Mills sector 
\cite{DKL, DFKZ1, L}: target-space duality of the modulus 
$T$ has a one-loop anomaly which is cancelled by a counterterm in
${\cal L}^{(1)}$, in a generalization to sigma-model anomalies of the
Green-Schwarz mechanism \cite{GSsigma}. The derivation of the 
complete counterterm requires a calculation to all
orders in the modulus $T$ \cite{DKL}. However, at the present 
stage of understanding, the M-theory approach should be regarded 
as a large-$T$ limit in which T-duality reduces to a 
shift symmetry in the imaginary part of $T$. 

Our goal in this section is to obtain some or all counterterms
in ${\cal L}^{(1)}$ associated with anomaly-cancellation in the low-energy 
description.  We are particularly interested in contributions 
to gauge kinetic terms, the so-called threshold corrections. 
And we want to formulate these terms using the
`M-theory multiplets' $V$, $V_T$ and $W$ corresponding to 
the surviving components of $G$, in contrast to the `heterotic
multiplets' $S$ (or $L$) and $T$. We begin by obtaining 
the relevant information from the case of heterotic $(2,2)$
symmetric orbifolds. 

\subsection{Information from symmetric (2,2) orbifolds}
\label{secorbi}

Retaining only the universal modulus $T$ and a linear dilaton 
multiplet $L$, the Wilson one-loop Lagrangian for heterotic
symmetric $(2,2)$ orbifolds includes a term\footnote{In this paragraph,
we use $c^a=1$ and $\hat L=L-2\sum_a\Omega^a$.} \cite{DFKZ1}
\beq
\label{22GS}
{\cal L}^{(1)} = -2\delta_{GS}\left[ \hat L\log(T+\ov T) \right]_D
+ \sum_a b^a [\log\eta(iT){\cal W}^a{\cal W}^a]_F\,,
\eeq
where $\delta_{GS}$ and $b^a$ are numbers depending on the orbifold 
and $\eta(iT)$ is the Dedekind function. Under $Sl(2,\Z)$ T-duality 
(\ref{modinv1}), the variation of ${\cal L}^{(1)}$ is
$$
\begin{array}{rcl}
\delta{\cal L}^{(1)} &=& 2\delta_{GS}\left[ \hat L
\{\log\varphi(T)+\log\ov\varphi(\ov T)\}\right]_D
+ {1\over2}b^a \left[ \log\varphi(T)\,
{\cal W}^a{\cal W}^a \right]_F \crbig
&=& {1\over2}(\delta_{GS}+b^a)\left[\log\varphi(T)\,
{\cal W}^a{\cal W}^a\right]_F,
\end{array}
$$
with $\varphi(T)=icT+d$. On the other hand, the triangle one-loop 
diagram for two gauge fields and one K\"ahler connection $-3\log(T+\ov T)$
is anomalous. Its variation is
$$
\delta\Delta = {1\over2} A^a \left[\log\varphi(T)\,
{\cal W}^a{\cal W}^a\right]_F,
$$
where $A^a$ is the chiral-anomaly coefficient, as obtained from the 
expression of the diagram. The anomaly cancels since one finds that
$b^a+\delta_{GS}+A^a=0$ 
for all factors in the gauge group (the index $a$). 
The one-loop correction to gauge kinetic terms obtained from the 
component expansion of ${\cal L}^{(1)}$ and from the triangle diagram 
reads
\beq
\label{thresh}
-{1\over4}F^a_{\mu\nu} F^{a\,\mu\nu}\left[ 
-(\delta_{GS}+A^a) \log(T+\ov T) + b^a\log|\eta(iT)|^4\right].
\eeq
It is modular invariant since the anomaly is cancelled,
and its value is controlled by the $F$-density contribution to
${\cal L}^{(1)}$, with coefficients $b^a$.

The Wilson Lagrangian depends on the coefficients $\delta_{GS}$ 
and $b^a$. But the information on gauge thresholds is in the 
numbers $b^a$. In general, the parameters of the Wilson Lagrangian 
computed at a non-trivial loop order are not of 
direct physical significance and this is here the case of $\delta_{GS}$
in the sector of gauge kinetic terms.

In the large-$T$ limit, T-duality reduces to
$\Im T\rightarrow\Im T+\hbox{constant}$, the K\"ahler connection
$-3\log(T+\ov T)$ is 
invariant and, strictly speaking, no anomaly survives to be cancelled. 
In addition, $\log|\eta(iT)|^4 \sim -{\pi\over3}(T+\ov T)$
dominates the logarithmic contributions. The threshold 
correction is then of the simple form
$$
-{1\over4}\sum_a\left[ -{\pi b^a\over3}(T+\ov T)
\right] F_{\mu\nu}^a F^{a\,\mu\nu},
$$
invariant under the imaginary shift symmetry of $T$. Its 
supersymmetrization is 
$$
-{\pi \over12}\sum_a b^a[T{\cal W}^a{\cal W}^a]_F = 
{2\pi \over3}\sum_a b^a[(T+\ov T)\Omega^a]_D\,. 
$$
As long as $T$ only is considered, there seems to be no way to identify
the $D$-density contribution to ${\cal L}^{(1)}$ in the large-$T$ limit. 
The introduction of the matter multiplet $M$ changes 
the picture. In the lowest order Wilson Lagrangian 
${\cal L}^{(0)}$, the K\"ahler connection $-3\log(T+\ov T)$ 
is modified to $-3\log(T+\ov T-2 \ov Me^A M)$. As a consequence,
the contribution with coefficient $A^a$ to the gauge threshold
(\ref{thresh}) involves the quantity $\log(T+\ov T-2\ov MM)$.
Then, either the one-loop term ${\cal L}^{(1)}$ is accordingly 
modified to $-2\delta_{GS}[\hat L\log(T+\ov T-2\ov M e^AM)]_D$
and the parameter $\delta_{GS}$ disappears from gauge thresholds,
or, less plausibly, this is not the case and $\delta_{GS}$ 
acquires a physical significance in $M$-dependent thresholds. 
In any case, since the holomorphic Dedekind function cannot 
depend on $\ov M e^A M$, a calculation of the correction 
${\cal L}^{(1)}$ to the Wilson Lagrangian to first order in 
$\ov M e^AM$ will give access to the $D$-density Green-Schwarz 
term and to the parameter $\delta_{GS}$.

From the point of view of M-theory on $O_7$ or heterotic strings on
Calabi-Yau, $\ov M e^A M$ arises from the ten-dimensional Chern-Simons terms,
at the same order as gauge kinetic terms. This indicates that the
$D$-density contribution to ${\cal L}^{(1)}$ should be visible in a reduction
of the ten-dimensional Green-Schwarz terms. 

\subsection{The case of M-theory on $O_7$}

Before embarking in a derivation of the anomaly-cancelling
terms from the low-energy limit of M-theory on $O_7$, 
we consider the problem at the level of four-dimensional 
supergravity only. 

In the large-$T$ limit, as discussed in the previous paragraph, the
$T$-dependent corrections to gauge kinetic terms are of the form
\beq
{1\over4}\sum_a \beta^a \left[ T{\cal W}^a{\cal W}^a\right]_F\,,
\eeq
with coefficients which are in principle calculable in heterotic 
strings. To rewrite them 
in terms of the `M-theory multiplets', we first note that
\beq
\label{gaugthr1}
{1\over 4}\sum_a\beta^a \left[ T{\cal W}^a{\cal W}^a \right]_F =
-2\Bigl[ (T+\ov T)\sum_a\beta^a\Omega^a \Bigr]_D.
\eeq
Since the field equation of the Lagrange multiplier
$L_T$ implies that $V_T=T+\ov T-2\ov Me^AM$, this expression 
can also be written
\beq
\label{gaugthr2}
-2\Bigl[ (V_T+2\ov Me^AM)\sum_a\beta^a\Omega^a\Bigr]_D.
\eeq
The right-hand side of Eq. (\ref{gaugthr1}) is gauge 
invariant because $\delta\Omega^a$ 
is a linear multiplet and therefore $\left[(T+\ov T)
\delta\Omega^a \right]_D$ is a derivative. To ensure 
gauge invariance of expression (\ref{gaugthr2}), we add 
the term $\left[L_T(V_T+2\ov Me^AM)\right]_D$ included 
in Lagrangian (\ref{matter3}). We obtain
\beq
\label{gaugthr3}
\Bigl[(L_T-2\sum_a\beta^a\Omega^a)(V_T+2\ov Me^AM)\Bigr]_D\,,
\eeq
which is gauge invariant if we postulate that
\beq
\label{LTvar}
\delta L_T = 2\sum_a\beta^a\delta\Omega^a.
\eeq

The correction (\ref{gaugthr1}) to the super-Yang-Mills 
Lagrangian is independent of the matter fields and has a 
holomorphic character (it can be seen as a correction to 
the holomorphic gauge kinetic function
$f_{ab}$). To enumerate possible matter-dependent contributions to 
gauge kinetic terms, we consider for simplicity 
a single matter multiplet $M$ transforming in some unspecified 
representation of the gauge group. 
The first candidate counterterm is a real density:
$$
\sum_a\gamma^a \Bigl[ \ov M e^A M \Omega^a\Bigr]_D. 
$$
Gauge invariance requires however its appearance in the
combination
$$
-2\delta\Bigl[\ov M e^A M(L-2\sum_ac^a\Omega^a)\Bigr]_D
+ 2\gamma\Bigl[\ov M e^A M(L_T-2\sum_a\beta^a\Omega^a)\Bigr]_D ,
$$
or
$$
-2\delta\Bigl[\ov M e^A M\, V\Bigr]_D
+ 2\gamma\Bigl[\ov M e^A M(L_T-2\sum_a\beta^a\Omega^a)\Bigr]_D ,
$$
using `M-theory multiplet' $V$. In the first counterterm, 
each gauge group factor contributes with weight $c^a$, 
as in lowest order terms. The second contribution can be 
combined with expression (\ref{gaugthr3}) into
\beq
\label{GS2}
\Bigl[(L_T-2\sum_a\beta^a\Omega^a)(V_T+2(1+\gamma)
\ov Me^AM)\Bigr]_D\,.
\eeq
We will also see below that the M-theory anomaly-cancelling 
terms generate a contribution of the form
\beq
\label{GSsuperpot}
\epsilon \Bigl[ V |\alpha M^3|^2 \Bigr]_D,
\eeq
involving the matter superpotential. Since
the factor $1+\gamma$ in expression (\ref{GS2}) can be 
eliminated by a rescaling of $M$, of $\delta$ 
and of the superpotential coupling constant $\alpha$, we 
may take $\gamma=0$ at our level of approximation, and the 
Wilson Lagrangian up to string one-loop order is expected 
to become
\beq
\label{matter4}
\begin{array}{rcl}
{\cal L} &=& 
\left[
-\Upsilon-(S+\ov S)(V+2\Omega) 
+\{U(W-\alpha M^3) +{\rm c.c.}\} \right.\crbig
&&+(L_T-2\sum_a\beta^a\Omega^a)(V_T+2\ov Me^AM)
\crbig
&&+ \left. V(\epsilon |\alpha M^3|^2 -2\delta\ov Me^AM)
\right]_D+[S_0^3W]_F\,.
\end{array}
\eeq
Notice that the contributions which correspond to string 
one-loop effects do not include any correction to the 
Einstein Lagrangian, which remains simply
$$
-{1\over2} \Upsilon eR, \qquad 
\Upsilon = (z_0\ov z_0 C_T)^{3\over2}(2C)^{-{1\over2}},
$$
and the Einstein frame condition remains $\Upsilon
=\kappa^{-2}$. This is expected since the gravitational 
constant in the heterotic string is not corrected at 
string one-loop order. 

From the general expression (\ref{matter4})
in which Bianchi identities are field equations for $S$, 
$L_T$ and $U$, we can derive various equivalent forms. For instance,  
solving for $S$, $L_T$ and $U$ leads to the version of the effective
supergravity in which the dilaton is described by a linear multiplet:
\beq
\label{linearvers}
\begin{array}{rcl}
{\cal L}_{\rm l} &=&
\Bigl[-(S_0\ov S_0)^{3/2}\left(T+\ov T-2\ov Me^AM\right)^{3/2}
(2\hat L)^{-1/2} \Bigr]_D
+ \Bigl[S_0^3(ih +\alpha M^3)\Bigr]_F \crbig
&&+\Bigl[\hat L\left(\epsilon|\alpha M^3|^2-2\delta\ov Me^AM
\right)\Bigr]_D 
+\Bigl[{1\over 4}T\sum_a\beta^a{\cal W}^a{\cal W}^a\Bigl]_F.
\end{array}
\eeq
The second line is a one-loop correction in the perturbative 
expansion of the heterotic string in which $\hat L$ is the string 
loop-counting field \cite{CFV}. Its $T$-dependent part corresponds 
to the Green-Schwarz counterterm found in \cite{DFKZ1} for 
symmetric heterotic orbifolds. Each of these one-loop corrections,
with coefficients $\epsilon$, $\delta$ and $\beta^a$,
is related to a well-defined counterterm which can be easily
identified in, for instance, the low-energy limit of M-theory on
$O_7$. The Green-Schwarz counterterms 
controlled by $\delta$ and $\epsilon$ are intrinsically real 
$D$-densities. They will appear as corrections to the K\"ahler 
potential, as matter-dependent `wave-function renormalisations',
in the dual version with a chiral dilaton multiplet.
On the other hand, the holomorphic $T$-dependent
terms are true threshold corrections. 

Alternatively, solving for $L_T$, $V$ and $U$ leads to the version
with a chiral dilaton multiplet:
\beq
\label{chiralvers}
{\cal L}_{\rm c} = 
-{3\over 2}\left[S_0\ov S_0\,e^{-K/3}\right]_D
+{1\over 4}\Bigl[\sum_a(c^aS+\beta^aT){\cal W}^a{\cal W}^a\Bigr]_F
+\Bigl[S_0^3(ih +\alpha M^3)\Bigr]_F,
\eeq
with the K\"ahler potential\footnote{The superfield K\"ahler 
potential
includes covariantization contributions $e^A$ which disappear in 
the bosonic expression used in component expansions.}
\beq
\label{fullK1}
\begin{array}{rcl}
K &=& -\log\left(S+\ov S+2\delta\ov Me^AM
-\epsilon|\alpha M^3|^2\right)  \crbig
&&-3\log\left(T+\ov T-2\ov Me^AM\right),
\end{array}
\eeq
and the gauge kinetic functions $f^a=c^aS+\beta^aT$. An ambiguity exists
however because one can perform a holomorphic redefinition of the 
two chiral multiplets. For instance,
\beq
\label{redef}
S = \tilde S - \delta T,
\eeq
leads to the equivalent K\"ahler potential
\beq
\label{fullK2}
\begin{array}{rcl}
K &=& -\log\left(\tilde S+\ov{\tilde S}-\delta(T+\ov T-2\ov Me^AM)
-\epsilon|\alpha M^3|^2\right)
\crbig
&&-3\log\left(T+\ov T-2\ov Me^AM\right),
\end{array}
\eeq
with gauge kinetic functions $f^a=c^a\tilde S+(\beta^a-c^a\delta)T$. 
The origin of this ambiguity at the level of `M-theory multiplets'
is interesting. Suppose that we add the counterterm 
$$
\Delta{\cal L} = 
A\left[ (V+2\Omega)(V_T+2\ov Me^AM) \right]_D
$$
to the fundamental Lagrangian (\ref{matter4}), with an arbitrary
constant $A$. The theory becomes then 
\beq
\label{matter5}
\begin{array}{rcl}
{\cal L}_A &=& 
\left[
-\Upsilon-(S+\ov S)(V+2\Omega) 
+\{U(W-\alpha M^3) +{\rm c.c.}\} \right.\crbig
&&+\left(L_T-2\sum_a(\beta^a-Ac^a)\Omega^a\right)
(V_T+2\ov Me^AM) \crbig
&&+ \left. V(\epsilon |\alpha M^3|^2 
+A V_T +2(A-\delta)\ov Me^AM) \right]_D+[S_0^3W]_F.
\end{array}
\eeq
It is gauge invariant provided the appropriate transformation 
of $L_T$ is postulated. 
We have apparently obtained a family of four-dimensional
supergravities, depending on a new parameter $A$. This is
however only true before solving for the Lagrange
multiplier multiplets. Firstly, solving for 
$S$ and $L_T$ leads to the space-time derivative $\Delta{\cal L}= 
A\left[L(T+\ov T) \right]_D$. The counterterm $\Delta{\cal L}$ is
then irrelevant in the version of the theory with a linear dilaton. 
Secondly, if we instead solve for $V$ and $L_T$, we obtain the
K\"ahler potential 
$$
\begin{array}{rcl}
K &=& -\log\left(S+\ov S -A(T+\ov T) +2\delta\ov Me^AM
-\epsilon|\alpha M^3|^2\right)  \crbig
&&-3\log\left(T+\ov T-2\ov Me^AM\right),
\end{array}
$$
and the gauge kinetic functions $f^a=c^aS+(\beta^a-Ac^a)T$. This 
theory is clearly related to Eq. (\ref{fullK1}) by the holomorphic
redefinition $S\rightarrow S-AT$, and the choice $A=\delta$
leads to theory (\ref{fullK2}). 

This discussion shows that the counterterm $\Delta{\cal L}$
is irrelevant in the four-dimensional effective supergravity,
that all values of $A$ lead to equivalent Lagrangians, with 
the same dynamical equations. Further information due, for
instance, to compactification of extra dimensions could however
appear more natural with a specific value of $A$, 
if one insists to use the version of the effective supergravity
 with a chiral dilaton multiplet. For instance, all 
corrections linear in $T$ appear as gauge thresholds with the
choice $A=0$. But one could as well use the version
with a linear dilaton which is free of ambiguities. 

In addition, the holomorphic redefinition (\ref{redef}) mixes 
terms of different orders in the string loop expansion. As a
consequence, the distinction between terms in ${\cal L}^{(0)}$ and
corrections in ${\cal L}^{(1)}$ becomes ambiguous in general in
the large $T$ limit. 

The values of the coefficients $\delta$ and $\epsilon$ 
can be inferred from a direct calculation of the $M$-dependent
anomaly-cancelling terms in M-theory on $O_7$. This is the subject 
of the next paragraph. Notice however that 
such a calculation only provides the terms of first order in
the matter multiplets $\ov Me^AM$ and $|\alpha M^3|^2$. 
To this order, expression (\ref{fullK1}) becomes
\beq
\begin{array}{rcl}
K&=&-\log(S+\ov S)
-3\log(T+\ov T-2\ov Me^AM) \crbig
&& -{1\over S+\ov S}\left[ 
2\delta\ov Me^AM-\epsilon|\alpha M^3|^2\right].
\end{array}
\eeq
The term with coefficient $\delta$ has been obtained in direct 
Calabi-Yau reductions of M-theory on $S^1/{\bf Z}_2$ 
(see for instance \cite{NOY,LOW1}\footnote{It has also been 
obtained, in a quite different context, by Itoyama and 
Leon \cite{IL}.}). The charged matter contribution 
with coefficient $\epsilon$ was not included 
in these analyses. 

The gauge contributions appearing in Eq. (\ref{matter4}) read
$$
-2\sum_a\left[\{c^a(S+\ov S)+\beta^a(V_T+2\ov 
Me^AM)\}\Omega^a\right]_D,
$$
so that the gauge coupling constants are given by
\beq
{1\over g_a^2}=c^a\Re s+{1\over 2}\beta^a\left(C_T+2
\ov MM \right).
\eeq
This expression becomes harmonic once the Bianchi 
identity imposing $C_T+2\ov MM=2\Re T$ has been used. 
Similarly, one obtains from Eq. (\ref{linearvers}) 
\beq
{1\over g_a^2}= {c^a\Upsilon\over4C}+\beta^a \Re T
-c^a\delta\,\ov MM +{c^a\epsilon\over2}|\alpha M^3|^2,
\eeq
with
$\Upsilon = 2(z_0\ov z_0)^{3/2}(\Re T-\ov MM)^{3/2} C^{-1/2}$.
This second form of the gauge couplings is never harmonic since it is
obtained from a theory with a linear dilaton multiplet. 
Both expressions do however agree since
the chiral-linear duality relation between $\Re s$ and $C$ is 
$$
\Re s = {\Upsilon\over4C}- \delta\ov MM
+{\epsilon\over2}|\alpha M^3|^2.
$$

\subsection{On the eleven-dimensional origin of the \\
anomaly-cancelling terms}\label{sec11d}

In ten dimensions, anomaly-cancelling terms for the 
$E_8\times E_8$ heterotic string are well-known. There are 
two terms. The first couples a gauge and Lorentz invariant 
eight-form $\hat X_8$ to the two-form field $\hat B$. The 
second one is proportional to $\int (\Omega_{3,1}+\Omega_{3,2}-
\Omega_{3{\rm L}})\wedge\hat X_7$, with $d\hat X_7= \hat X_8$. 

The anomaly-cancelling terms arising from M-theory on $S^1/
{\bf Z}_2$ \cite{HW1, HW2} have recently been precisely 
computed \cite{BDS}. They arise from the following action terms
in eleven dimensions \cite{HW1, HW2, GX7}:
\beq
\label{11dGS1}
-{\lambda^2\over(4\pi)^2\kappa^2_{11}}\int G\wedge X_7
-{1\over12\kappa^2_{11}}\int C\wedge G\wedge G,
\eeq
where
\beq
\label{X7is}
X_7 = {1\over12(4\pi)^3}\left( {1\over2}\Omega_{7{\rm L}}
-{1\over8}({\rm tr}\,R^2)\Omega_{3{\rm L}}  \right),
\eeq
$\kappa_{11}$ is the eleven-dimensional gravitational constant
and $\lambda$ is the gauge coupling constant on both ten-dimensional
${\bf Z}_2$-fixed planes, as defined by the gauge action 
(\ref{Sgauge}). The Chern-Simons forms are defined by
$$
d\Omega_{7{\rm L}} = {\rm tr}\, R^4, \qquad
d\Omega_{3{\rm L}} = {\rm tr}\, R^2, \qquad
d\Omega_{3,i} = {\rm tr}\, F_i^2, \qquad i=1,2.
$$
Cancellation of gauge and gravitational 
anomalies and coherence of the reduction to ten dimensions 
impose $\lambda^6 = (4\pi)^5\kappa^4_{11}/12$ 
\cite{HW1, HW2, BDS}. This condition relates the gauge 
coupling $\lambda$ and the $S^1$ radius. Solving the Bianchi 
identity verified by the four-form field $G$ on $S^1/{\bf Z}_2$ 
and extracting the zero modes leads to the following 
Green-Schwarz terms 
\beq
\label{11dGS2}
\begin{array}{rcl}
S_{GS} &=& -{1\over48\pi}\displaystyle{\int_{M_{10}}}
\hat B\wedge\left[ (I_{4,1})^2 + (I_{4,2})^2
-I_{4,1}I_{4,2} + {1\over(4\pi)^4}\left( 
{1\over2}{\rm tr}\, R^4 -{1\over8}({\rm tr}\, R^2)^2 
\right)\right] 
\crbig
&&
-{1\over(4\pi)^2}\displaystyle{\int_{M_{10}}} (\Omega_{3,1}
+\Omega_{3,2}-\Omega_{3{\rm L}})\wedge X_7.
\end{array}
\eeq
In this expression,
\beq
\label{11dGS3}
I_{4,i} = {1\over(4\pi)^2}\left( {\rm tr}\, F_i^2
-{1\over2}{\rm tr}\, R^2 \right), \qquad i=1,2,
\eeq
and\footnote{Recall that $x^4$ is the $S^1$ coordinate.}
\beq
\label{lambdais}
\hat B_{AB} = {\lambda^2\over\kappa^2_{11}}\int_{S^1}dx^4\, 
C_{AB4}\,.
\eeq
As usual, the  two-form field $\hat B$ couples to $\hat X_8$.
But the second term has a particular structure: $\hat X_7$
is replaced by the purely gravitational seven-form (\ref{X7is}): the
anomaly-cancelling terms derived from M-theory differ
from the standard expression of the heterotic string by a 
well-defined local counterterm, as permitted by the descent 
equations \cite{BDS}. 

The Calabi-Yau compactification of $\hat B_{AB}$ leads to two
zero modes of the bulk fields,
$$
\kappa^2\hat B_{\mu\nu} = {\lambda^2\over V_6} C_{\mu\nu4},
\qquad\qquad
\kappa^2\hat B_{i\ov j} = {\lambda^2\over V_6} C_{i\ov j4},
$$
since
${\lambda^2\over\kappa^2_{11}} = {1\over\kappa^2}\,
{\lambda^2\over V_6}\,{1\over V_1}$. By Eqs. (\ref{CJScompident}), 
these states are related to our four-dimensional 
bulk multiplets $L$ (or $S$) and $T$ by 
\beq
\label{bulkbulk}
\hat B_{\mu\nu} = 4b_{\mu\nu}, 
\qquad\qquad
\kappa^2\hat B_{i\ov j} = i\Im T\,\delta_{i\ov j}.
\eeq
The number $\lambda^2/V_6$ disappears in this identification: it
does not play any role in the four-dimensional effective theory.

Our task is then to compute the reduction of $S_{GS}$ on
$M_4\times$(Calabi-Yau). Since we restrict ourselves to
contributions with at most two derivatives, we only need the 
reduction of
$$
\Delta = -{1\over48\pi}\displaystyle{\int_{M_{10}}}
\hat B\wedge\left[(I_{4,1})^2 + (I_{4,2})^2
-I_{4,1}I_{4,2} \right].
$$
The global definition of the four-form field $G$ 
(`cohomology condition') implies \cite{W}
$$
\langle I_{4,1}\rangle = -\langle I_{4,2}\rangle
$$
for the Calabi-Yau background, which is a $(2,2)$ form. 
The counterterm $\Delta$ becomes then
$$
\begin{array}{rcl}
\Delta &=& -{1\over 4(4\pi)^3}\displaystyle{\int_{M_{10}}}
\hat B\wedge \langle I_{4,1} \rangle\wedge \left(
{\rm tr}\, F_1^2 - {\rm tr}\, F_2^2 \right) \crbig
&&-{1\over 12(4\pi)^5}\displaystyle{\int_{M_{10}}}
\hat B\wedge\left[({\rm tr}\, F_1^2)^2 + ({\rm tr}\, F_2^2)^2
- ({\rm tr}\, F_1^2) ({\rm tr}\, F_2^2) \right] + \ldots,
\end{array}
$$
where gravitational contributions with more than two derivatives 
are omitted.
Notice that the two $E_8$ factors contribute with opposite signs
in the background-dependent term. 

The standard embedding is defined by $\langle {\rm tr}\, F_2^2
\rangle =0$, while $\langle {\rm tr}\, F_1^2\rangle$ is in the
$SU(3)$ direction of the maximal embedding $E_6\times SU(3)
\subset E_8$. As a consequence, 
$$
\langle {\rm tr}\, F_1^2\rangle =\langle {\rm tr}\, R^2\rangle
= 2(4\pi)^2 \langle I_{4,1}\rangle,
$$
which in turn leads to
\beq
\label{Delta}
\begin{array}{rcl}
\Delta &=& -{1\over 8(4\pi)^5} \displaystyle{\int_{M_{10}}}
\hat B\wedge\langle{\rm tr}\, R^2\rangle\wedge (
{\rm tr}\, F_1^2 -{\rm tr}\, F_2^2 ) \crbig
&&-{1\over 12(4\pi)^5}\displaystyle{\int_{M_{10}}}
\hat B\wedge\Bigl[({\rm tr}\, F_1^2)^2+({\rm tr}\, 
F_2^2)^2 -({\rm tr}\, F_1^2)({\rm tr}\, F_2^2)\Bigr]+\ldots.
\end{array}
\eeq

To derive the zero modes of ${\rm tr}\, F_1^2$ and
${\rm tr}\, F_2^2$, it is simpler to consider the Chern-Simons 
forms, using the relation
$$
({\rm tr}\, F_i^2)_{ABCD} = 
4\,\partial_{[A} (\Omega_{3,i})_{BCD]}, \qquad i=1,2.
$$
In the standard embedding, the unbroken $E_8$ group generates
a $N=1$ super-Yang-Mills multiplet only. The only massless mode is then
$$
({\rm tr}\, F_2^2)_{ABCD}  \,\,\longrightarrow\,\,
({\rm tr}\, F_{E_8}^2)_{\mu\nu\rho\sigma}, \qquad
(\Omega_{3,2})_{ABC} \,\,\longrightarrow\,\,
(\Omega_{E_8})_{\mu\nu\rho}. 
$$
Clearly, these massless modes of ${\rm tr}\, F_2^2$ and $\Omega_{3,2}$
are respectively components of the four-dimensional $N=1$ multiplets 
${\cal W}^2{\cal W}^2$ and $\Omega^2$ used earlier. More precisely:
\beq
\label{compgauge}
\begin{array}{rcl}
[{\cal W}^2{\cal W}^2]_{\rm f-component} &=&
-{1\over2}\,{\rm tr}\, F_{E_8\,\mu\nu}F_{E_8}^{\mu\nu}
-{i\over4e}\epsilon^{\mu\nu\rho\sigma}{\rm tr}
\,F_{E_8\,\mu\nu}F_{E_8\,\rho\sigma}+\ldots, \crbig
[\Omega^2]_{\rm d-component } &=&
{1\over16}\,{\rm tr}\, F_{E_8\,\mu\nu}F_{E_8}^{\mu\nu} + \ldots,
\crbig
[\Omega^2]_{\rm vector \,\, component } &=&
{1\over8e}\epsilon^{\mu\nu\rho\sigma}(\Omega_{E_8})_{\nu\rho\sigma}
+\ldots
\end{array}
\eeq
The gauge fields of the $E_8$ group broken into $E_6$ generate 
$E_6$ gauge fields and the chiral matter multiplet $M$, 
transforming in representation ${\bf 27}$. Accordingly, 
the massless modes of ${\rm tr}\, F_1^2$ are:
$$
\begin{array}{rcccl}
({\rm tr}\, F_1^2)_{ABCD}  &\,\,\longrightarrow\,\,& 
({\rm tr}\, F_{E_6}^2)_{\mu\nu\rho\sigma},&& \crbig
&\,\,\longrightarrow\,\,& ({\rm tr}\, F_1^2)_{\mu\nu i\ov j}
&=& 2\,\partial_{[\mu}(\Omega_{3,1})_{\nu] i\ov j}\,, 
\crbig
&\,\,\longrightarrow\,\,& ({\rm tr}\, F_1^2)_{\mu ijk}
&=&\partial_\mu(\Omega_{3,1})_{ijk}\,, \crbig
&\,\,\longrightarrow\,\,& ({\rm tr}\, F_1^2)_{\mu\ov{ijk}}
&=&\partial_\mu(\Omega_{3,1})_{\ov{ijk}}\,.
\end{array}
$$
While as before $(\Omega_{3,1})_{ABC} \,\rightarrow\,
(\Omega_{E_6})_{\mu\nu\rho}$, 
the other components of $\Omega_{3,1}$ involve the scalar component 
of the matter multiplet $M$ and require more care since
we have already precisely defined the four-dimensional
field $M$ by its coupling to the bulk fields. To obtain
the correct relations, a detour is helpful. 

As already explained, the gauge kinetic action (\ref{Sgauge})
also generates the four-dimen\-sio\-nal kinetic terms
for the matter multiplet $M$. In the four-dimensional effective 
Lagrangian, these contributions arise as the highest component of 
the `matter Chern-Simons multiplet' $\ov M e^AM$, which includes
$\,-2(D_\mu\ov M)(D^\mu M)\,$. This multiplet also contains in its 
vector component the matter Chern-Simons form
$$
\Omega^M_\mu = i\ov M(D_\mu M) - i(D_\mu\ov M) M.
$$
This is completely similar to the gauge Chern-Simons multiplet 
which includes gauge kinetic terms in its d-component and 
$(\Omega_3)_{\mu\nu\rho}$ in its vector component, as indicated 
by expressions (\ref{compgauge}).
A direct computation of the relation between kinetic terms 
due to the action (\ref{Sgauge}), and the highest component 
of $\ov Me^AM$ delivers then the relation between 
$(\Omega_{3,1})_{\mu i\ov j}$ and this multiplet, by 
four-dimensional supersymmetry. A similar operation gives 
the relation between $(\Omega_{3,1})_{ijk}$ and the 
superpotential multiplet $\alpha M^3$. The relations are
\beq
\label{Omident}
\begin{array}{rcl}
(\Omega_{3,1})_{\mu i\ov j} &=&
{1\over6\kappa^2}\left[(D_\mu\ov M)M - \ov M(D_\mu M)\right]
\delta_{i\ov j} \,\,=\,\, {i\over6\kappa^2}
\delta_{i\ov j}\,[\ov M e^A M]_{\rm vector \,\, component}\,,
\crbig
(\Omega_{3,1})_{ijk} &=& 
{1\over3\kappa^3}\alpha\ov M^3\epsilon_{ijk}
\,\,=\,\, {1\over3\kappa^3}\,\epsilon_{ijk}\,
[\alpha\ov M^3]_{\rm scalar \,\, component},
\end{array}
\eeq
together with the last equation (\ref{compgauge}) which applies 
to all gauge Chern-Simons forms. 

With the complete identification of the massless modes of the 
Chern-Simons three-forms, we are equipped for 
translating the Calabi-Yau reduction of the Green-Schwarz
counterterm (\ref{Delta}) in a four-dimensional supergravity 
density formula. After 
straightforward manipulations of Eq. (\ref{Delta}), we obtain
\beq
\label{final1}
\begin{array}{rcl}
{\cal L}_{GS} &=&
-{1\over192(4\pi)^5} I\, \epsilon^{\mu\nu\rho\sigma}
{\lambda^2\over V_6}(\partial_\mu a)(\Omega_{E_6}-\Omega_{E_8})
_{\nu\rho\sigma} \crbig
&&+{i\over384(4\pi)^5\kappa^2} I\, \epsilon^{\mu\nu\rho\sigma}
{\lambda^2\over V_6}(\partial_\mu C_{\nu\rho4})(MD_\sigma\ov M
-\ov MD_\sigma M) \crbig
&&+{i\alpha^2\over 216(4\pi)^5}{V_6\over\kappa^6}
\epsilon^{\mu\nu\rho\sigma}
{\lambda^2\over\kappa^2 V_6}(\partial_\mu C_{\nu\rho4})
\left[\ov M^3(\partial_\sigma M^3) 
-(\partial_\sigma\ov M^3)M^3 \right]+\,\ldots \crbig
&=&
{1\over384(4\pi)^5} I\, \epsilon^{\mu\nu\rho\sigma}
T_\mu (\Omega_{E_6}-\Omega_{E_8})_{\nu\rho\sigma} \crbig
&&-{i\over96(4\pi)^5} I\, \epsilon^{\mu\nu\rho\sigma}
(\partial_\nu b_{\rho\sigma})(MD_\mu\ov M-\ov MD_\mu M) \crbig
&&-{i\alpha^2\over 54(4\pi)^5} {V_6\over\kappa^6} 
\epsilon^{\mu\nu\rho\sigma}(\partial_\nu b_{\rho\sigma})
\left[\ov M^3(\partial_\mu M^3) 
-(\partial_\mu\ov M^3)M^3 \right]+\,\ldots
\end{array}
\eeq
where the dots indicate the terms required by
$N=1$ supersymmetry and $I$ is the dimensionless integral
$$
I = \kappa^{-2}\int_{CY} dV_6 \, \delta_{i\ov i}\epsilon^{ijk}
\epsilon^{\ov{ijk}}\langle{\rm tr}\, R^2\rangle_{jk\ov{jk}},
$$
in terms of the $(1,1)$ (the metric tensor), $(3,0)$ and $(0,3)$ 
Calabi-Yau tensors and the background $\langle{\rm tr}\, R^2\rangle$.

Using Eqs. (\ref{Omident}), we can write the Green-Schwarz 
counterterm in superfield form as:
\beq
\label{final2}
\begin{array}{rcl}
{\cal L}_{GS} &=&
-{I\over 48(4\pi)^5}\left[(V_T+2\ov Me^AM)
(\Omega^1-\Omega^2)\right]_D
-{I\over 48(4\pi)^5}\left[V\ov Me^AM\right]_D \crbig
&&
+{1\over 27(4\pi)^5}{V_6\over \kappa^6}\left[V|\alpha M^3|^2
\right]_D.
\end{array}
\eeq
Comparing the three terms of ${\cal L}_{GS}$ with the corresponding 
parts of (\ref{matter4}), we can express the coefficients $\beta^a$, 
$\delta$ and $\epsilon$ in terms of the Calabi-Yau dependent integral 
$I$:
\beq
\delta = \beta^1=-\beta^2={I\over 96(4\pi)^5}~~,~~
\epsilon={1\over 27(4\pi)^5}{V_6\over \kappa^6}\,.
\eeq
The calculation predicts then $\beta^1=-\beta^2=\delta$, a result
already obtained in references \cite{NOY, LOW1}, for instance.
Once again, we stress that we have obtained next-order corrections
to the effective Wilson Lagrangian. As argued earlier in paragraph
\ref{secorbi}, while we expect $\beta^1$ and $\beta^2$ to have 
physical significance as coefficients of the modulus-dependent 
threshold corrections, the parameter $\delta$ is not necessarily 
a physical quantity. To decide of its relevance, a calculation at
the same order of threshold corrections in the effective action 
should be performed, but this computation requires a detailed 
knowledge of the charged matter spectrum and couplings. 

\section{Conclusions}\label{secfinal}

In this paper, we have deduced the structure of the
four-dimensional $N=1$ effective (wilsonnian) supergravity 
describing the universal massless sector of M-theory compactified 
on $(\hbox{CY})_3\times S^1/{\bf Z}_2$. The theory depends on three 
categories of multiplets: `M-theory multiplets' $V$, $V_T$ and $W$ 
describe the degrees of freedom of the M-theory four-index tensor, 
`source multiplets' 
$\Omega^a$, $\ov Me^A M$ and $\alpha M^3$ are related to the source
terms in M-theory Bianchi identities and `Lagrange multiplets'
$S$, $L_T$ and $U$ impose by their field equations the Bianchi
identities. In addition, the multipliers $S$ and $L_T$ generate 
the four-dimensional axion-tensor duality which is known to be
an important ingredient of the formulation of the string dilaton 
beyond the lowest order. 

An effective supergravity similar to expression (\ref{matter4}) is
in principle valid for generic compactifications of M-theory
with unbroken $N=1$ supersymmetry in four dimensions. One needs to
identify the appropriate supermultiplets appearing as sources in the
Bianchi identities generated by $S$, $L_T$ and $U$. Only the values
of coefficients like $\beta^a$, $\delta$ and $\epsilon$ depend on 
the detailed geometry of the compact space.
This method is especially useful in deriving contributions to the
effective Lagrangian due to non-perturbative
states like M-theory five-brane degrees of freedom or condensates.
Some of these more general $N=1$ vacua will be studied in a
forthcoming publication \cite{DS2}.      

\vspace{.5cm}
\begin{center}{\bf Acknowledgements}\end{center}
\noindent
The authors have benefited from discussions with A. Bilal, C. 
Kounnas, A. Lukas, D. L\"ust and B. Ovrut. This research 
was supported in part by the European Union under the TMR 
contract ERBFMRX-CT96-0045, the Swiss National Science Foundation 
and the Swiss Office for Education and Science.
\vspace{.5cm}

\section*{Appendix: Notations and conventions}
\label{appnotation}
\renewcommand{\theequation}{A.\arabic{equation}}
\setcounter{equation}{0}

{\it Metrics and coordinates:}

\noindent
The space-time metric has signature $(-,+,+,\ldots,+).$

\noindent
For coordinates, our notation is: 

\begin{tabular} {lll}
$D=11$ curved space-time: & $x^M$ & $M=0,\ldots,10$ \\
$D=10$ curved space-time: & $x^A$ & \\
$D=4$ curved space-time: & $x^\mu$ & $\mu =0,1,2,3$ \\
$S_1/{\bf Z}_2$ direction: & $x^4$ & \\
Calabi-Yau directions: & $x^a$ & $a=5,\ldots,10$  \\
Calabi-Yau complex (K\"ahler) coordinates: \hspace{5mm}& 
$z^i$, $\ov z^{\ov i}$ & $i=1,2,3$ 
\end{tabular}

\noindent
For reduction purposes, we simply use
$$
z^l = {1 \over \sqrt{2}}\left(x^l+ix^{l+3}\right)\,,
\qquad\qquad 
\ov z^{\ov l} = {1 \over \sqrt{2}}\left(x^l-ix^{l+3}\right)\,,
\qquad\qquad l=1,2,3.
$$ 
$\epsilon_{ijk}$ is the $SU(3)$--invariant Calabi-Yau tensor 
such that $\epsilon_{123}=\epsilon_{\ov{123}}=1$.

\medskip
\noindent{\it Antisymmetric tensors:}

\noindent
Antisymmetrization of $n$ indices has unit weight:
$$
A_{[M_1\ldots M_n]} = {1\over n!}\Bigl(A_{M_1\ldots M_n} \pm 
(n!-1){\rm \,\,permutations}\Bigr).
$$

\medskip
\noindent{\it Differential forms}

\noindent
For a $p$--index antisymmetric tensor, we define
$$
A^{(p)} = {1\over p!}A_{M_1\ldots M_p}\, 
dx^{M_1}\wedge\ldots\wedge dx^{M_p}. 
$$
Then, 
$$
\begin{array}{l}
A^{(p)}\wedge B^{(q)} = {1\over p!q!}A_{M_1\ldots M_p}
B_{M_{p+1}\ldots M_{p+q}} \,dx^{M_1}\wedge\ldots\wedge 
dx^{M_{p+q}} = C^{(p+q)}, \crbig
C_{M_1\ldots M_{p+q}} = {(p+q)!\over p!q!}A_{[M_1\ldots M_p} 
B_{M_{p+1}\ldots M_{p+q}]}.
\end{array}
$$
The exterior derivative is $d=\partial_M\,dx^M$. 
The curl $F^{(p+1)} = dA^{(p)}$ of a $p$-form reads then
$$
\begin{array}{rcl}
dA^{(p)} &=& {1\over p!} (\partial_MA_{N_1\ldots N_p})\,
dx^M\wedge dx^{N_1}\wedge\ldots\wedge dx^{N_p}  \crbig
&=& {1\over (p+1)!} F_{M_1\ldots M_{p+1}} \, dx^{M_1}\wedge\ldots
\wedge dx^{M_{p+1}}, \crbig
F_{M_1\ldots M_{p+1}} &=& (p+1)\, \partial_{[M_1}A_{M_2\ldots 
M_{p+1}]} \crbig
&=& \partial_{M_1}A_{M_2\ldots M_{p+1}} \pm \,\, p
{\rm \,\,\, cyclic\,\,permutations}\,.
\end{array}
$$
The volume form in $D$ space-time dimensions is
$dx^{M_1}\wedge\ldots\wedge dx^{M_D} = \epsilon^{M_1\ldots 
M_D}\,d^Dx$.
\\ \noindent We use analogous conventions for forms in four
space-time dimensions.

\end{document}